\documentclass[tex,twocolumn,epjc3]{svjour3}
\smartqed  

\RequirePackage{graphicx}
\RequirePackage{ulem}
\RequirePackage{epsfig,multicol,bbm,euscript,array, bbm, euscript}
\RequirePackage{flushend}
\RequirePackage{amssymb}
\RequirePackage{amsmath}
\RequirePackage{mathrsfs}
\RequirePackage{amsfonts}
\RequirePackage{latexsym}

\RequirePackage{fix-cm}

\journalname{Eur. Phys. J. C}
\begin{document}

\title{Covariant holography of a tachyonic accelerating universe
}

\author{Alberto Rozas-Fern\'{a}ndez\thanksref{e1,addr1,addr2}
}

\thankstext{e1}{e-mail: a.rozas@iff.csic.es}

\institute{Instituto de F\'{\i}sica Fundamental,
Consejo Superior de Investigaciones Cient\'{\i}ficas, Serrano 121,
28006 Madrid, Spain \label{addr1}
           \and
Institute of Cosmology \& Gravitation, University of Portsmouth,
Dennis Sciama Building, Portsmouth, PO1 3FX, UK \label{addr2}
}

\date{Received: 15 May 2014 / Accepted: 31 July 2014}

\maketitle

\begin{abstract}
We apply the holographic principle to a flat dark energy dominated Friedmann-Robertson-Walker spacetime
filled with a tachyon scalar field with
 constant equation of state $w=p/\rho$, both for $w>-1$ and $w<-1$. By using a
  geometrical covariant procedure, which allows the construction of holographic
   hypersurfaces, we have obtained for each case the position of the preferred screen
    and have then compared these with those obtained by using the holographic dark
     energy model with the future event horizon as the infrared cutoff. In the phantom scenario, one
     of the two obtained holographic screens is placed on the big rip hypersurface, both for
      the covariant holographic formalism and the holographic phantom model. It is also analyzed
       whether the existence of these preferred screens allows a mathematically consistent formulation of fundamental
        theories based on the existence of a S matrix at infinite distances.

\end{abstract}

\section{Introduction}
\label{intro}
The holographic principle was put forward in 1993 \cite{'tHooft:1993gx,Susskind:1994vu} and asserts that all of the information contained in some region of space can be represented as a hologram, a theory located on the boundary of that region. This theory should contain at most one degree of freedom per Planck area. Since then,
 the holographic principle has been fruitfully developed, as in its most well known implementation, the AdS/CFT correspondence \cite{Maldacena:1997re}, as well as in its connection with M-theory \cite{Horava:1997dd}. A
cosmological version of the holographic principle was proposed
in \cite{Fischler:1998st}.

We shall study in this paper how the holographic principle applies to an accelerating universe filled with a tachyon scalar field with a constant equation of state (EoS) $w=p/\rho$, for both the spacetimes  of the region $-1<w<-1/3$ and the phantom domain $w<-1$. In order to do so, we shall make use of a covariant procedure \cite{Bousso:2002ju} and the results will be confronted with those given by the holographic dark energy model with the future event horizon as the infrared cutoff \cite{Li:2004rb}. Our analysis is relevant in at least two aspects . On the one hand, dark energy should contain a large amount of the relevant degrees of freedom and hence, in order to constrain the EoS for dark energy, it is important to investigate whether such degrees of freedom are projected on the same boundary surfaces as those characterising the remaining non-vacuum energy. On the other hand, a universe with constant EoS $w$ that accelerates indefinitely will exhibit a future event horizon \cite{He:2001za} (see, however, \cite{GonzalezDiaz:2001ce}), presenting a challenge for string theories because it is not possible to construct a conventional S-matrix as the local observer inside his horizon is not able to isolate particles to be scattered. The emergence of an event horizon at the future, which would behave as a holographic screen, would aggravate this problem.

This paper can be outlined as follows. Sec.\ \ref{sec:tachyonDE} contains the spacetime of a flat Friedmann-Robertson-Wal-ker (FRW) universe filled
with a tachyon scalar field in the region $w>-1$. In Sec.\ \ref{sec:covtachyonDE} a covariant formalism is used to
derive the holographic preferred screens that correspond to
the spacetime presented in Sec.\ \ref{sec:tachyonDE}. In Sec.\ \ref{sec:phtachyonDE} we discuss
the covariant holography of a flat tachyonic phantom energy scenario. In Sec.\ \ref{sec:HDE},
the dark and phantom holographic dark energy models are constructed for a flat geometry in order to insert
holographic screens in terms of the future event
horizon \cite{Li:2004rb} or the horizon at the big rip \cite{GonzalezDiaz:2005sh}. The conclusions are drawn in Sec.\ \ref{sec:concl}.

\section{The spacetime of a universe filled with tachyonic dark energy}\label{sec:tachyonDE}

The explanation of dark energy is a central preoccupation of present-day cosmology.
In the $\Lambda$CDM paradigm, in which the cosmological constant accounts for the acceleration of the universe, the
universe would asymptotically tend to the de Sitter space-time whose
holographic properties have already been studied in some depth \cite{Karch:2003em,Alishahiha:2005dj}.
However, the dark energy could perfectly be dynamical in nature, even favored over the cosmological constant \cite{Samushia:2012iq}.

If we
 consider the tachyon as a dark energy candidate, the spacetime structure that results presents some holographic properties that have not been
 considered yet and that deserve our attention.

The fact that the tachyon can act as a source of dark energy with
different potential forms have been widely discussed in the
literature \cite{Bagla:2002yn,Padmanabhan:2002cp,Gibbons:2002md,Frolov:2002rr,Shao:2007zv,Calcagni:2006ge,Copeland:2004hq}. The tachyon can
be described by an effective field theory corresponding to a
tachyon condensate in a certain class of string theories with the
following effective action \cite{Bergshoeff:2000dq,Sen:2000kd,Sen:2004nf}

\begin{equation}\label{eq21}
S=\int d^4x\sqrt{-g}\left[\frac{R}{16\pi
G}-V(\phi)\sqrt{1+g^{\mu\nu}\partial_{\mu}\phi\partial_{\nu}\phi}\right],
\end{equation}

where $V(\phi)$ is the tachyon potential and $R$ the Ricci scalar.
The physics of tachyon condensation is described by the above
action for all values of $\phi$ provided the string coupling and
the second derivative of $\phi$ are small.

 The corresponding
energy-momentum tensor of the tachyon field has the form

\begin{equation}\label{eq22}
T_{\mu\nu}=\frac{V(\phi)\partial_{\mu}\phi\partial_{\nu}\phi}{\sqrt{1+g^{\alpha\beta}\partial_{\alpha}\phi\partial_{\beta}\phi}}-g_{\mu\nu}V(\phi)\sqrt{1+g^{\alpha\beta}\partial_{\alpha}\phi\partial_{\beta}\phi}\;.
\end{equation}

Let us now consider a spatially flat FRW spacetime with line element
\begin{equation}\label{metric}
ds^{2}=-dt^{2}+a(t)^{2}d\vec{x}^{2}\;,
\end{equation} in which $a(t)$ is the scale factor. The Friedmann equations then read

\begin{equation}\label{Fried1}
H^{2} \equiv \left(\frac{\dot{a}}{a}\right)^{2} =\frac{8\pi G \rho_{t}}{3}
\end{equation}

\begin{equation}\label{Fried2}
\frac{\ddot{a}}{a}=-\frac{4 \pi G(\rho_{t}+3p_{t})}{3}
\end{equation} where the energy density $\rho_{t}$ and the
pressure $p_{t}$ are given by
\begin{equation}\label{eq23}
\rho_t=-T_{0}{}^{0}=\frac{V(\phi)}{\sqrt{1-\dot{\phi}^2}}\;,
\end{equation}
\begin{equation}\label{eq24}
p_t=T_{i}{}^{i}=-V(\phi)\sqrt{1-\dot{\phi}^2}\;,
\end{equation} and the dot
stands for the derivative with respect to cosmic time.

From Eqs. (\ref{eq23}) and (\ref{eq24}) we obtain the tachyon
EoS parameter
\begin{equation}\label{eq25}
w=\frac{p_{t}}{\rho_{t}}=\dot{\phi}^2-1
\end{equation} and we shall consider in what follows that $w>-1$.
We shall
also restrict ourselves to consider a description of the current cosmic situation where
it is assumed that the tachyon component largely dominates and therefore we
shall disregard the non-relativistic and relativistic components of
the matter density and pressure.

If we assume a linear time-dependence of
the tachyon field $\phi$ and hence constancy of the parameter $w$, then the general expression for $a(t)$ can be written as

\begin{equation}\label{scfact}
a(t)=\left[a_{0}^{3(1+w)/2}+\frac{3}{2}(1+w)(t-t_{0})\right]^{2/[3(1+w)]}
\end{equation} where $a_{0}$ is the initial value of the scale factor at the initial time $t_{0}>0$. This solution describes an
accelerating universe in the interval $-1<w<-1/3$.
In order to facilitate the study of the holographic properties of the
tachyonic spacetime, it is best to express solution
(\ref{scfact}) in terms of the conformal time

\begin{equation}\label{conftime}
\eta=\int\frac{dt}{a(t)}=\frac{2a^{(1+3w)/2}}{1+3w}\;.
\end{equation}
Note that $-\infty<\eta
< 0$ for $w < -1/3$. Therefore, the scale factor in terms of the conformal time now reads
\begin{equation}
a(\eta)=\left[\frac{(1+3w)\eta}{2}\right]^{2/(1+3w)}\;.
\end{equation}

\section{Covariant holography in a tachyonic accelerating universe}\label{sec:covtachyonDE}
In this section we shall carry out the study of the holographic properties of the space-time presented in Sec.\ \ref{sec:tachyonDE}, following the covariant formalism developed in \cite{Bousso:2002ju} for general space-times. We shall start first by drawing the Penrose diagram for
our tachyonic asymptotic spacetime and then we shall construct the embedded holographic hypersurfaces (screens), which are surfaces on which the information in the space-time bulk can be encoded at less than
 one bit per Planck area \cite{'tHooft:1993gx,Susskind:1994vu}. In order to construct screens, we
must slice the spacetime into a family of light-cones centred at $r=0$ that can be parameterised by time and then identify in which direction to project among the two inequivalent null projections, which go along
past or future-directed light cones.

The Penrose diagram is constructed by mapping our FRW space-time on a part of the Einstein static universe \cite{Hawking:1973uf}, whose causal structure is that of an infinite cylinder $R \times S^{3}$, and determining the regions of it that are conformal to our FRW space-time. In the resulting Penrose diagram, every point represents a $S^{2}$ sphere and each diagonal line represents a light-cone. The two inequivalent null slicings can be represented by the ascending and descending families of diagonal lines. We then proceed to identify the apparent horizons, which are defined geometrically as the spheres (hypersurfaces) at which at least one pair (past or future) of orthogonal null congruences has zero expansion. These horizons will divide the space-time into normal, trapped and anti-trapped regions \cite{Bousso:2002ju,Hawking:1973uf}.

We shall finally determine the preferred and
optimal (if any) screen hypersurfaces which are going to encode
all the information in the universe. A preferred screen
is a surface in which the expansion of all
projected null hypersurfaces becomes zero at every point \cite{Bousso:2002ju}. If the expansions of
both independent pairs of orthogonal families of light-rays vanish
on one of the preferred screens, it becomes an optimal screen
\cite{Bousso:2002ju}.

A flat FRW spacetime is described, in terms of the conformal time $\eta$, by a metric of the form

\begin{equation}\label{confmetric}
ds^2= a(\eta)^2\left(-d\eta^2 +dr^2 +r^2 d\Omega_2^2\right)\;,
\end{equation} where $0 < r < \infty$, and $d\Omega_2^2 = d\theta^2 +\sin^2\theta
d\phi^2$ is the metric on the unit $S^{2}$ sphere, with $0<\theta
<\pi$ and $0 <\phi <2\pi$. This metric can be reduced to a more convenient form \cite{Hawking:1973uf} by defining
some new coordinates, $p$ and $q$, such that
$t'=p+q$ and $r'=p-q$. This allows the metric (\ref{confmetric}) to be expressed in a form
which is conformal to that of Minkowski space in spherical
coordinates, and hence locally identical to that of the Einstein static universe

\begin{eqnarray}\label{Einsteinst}
&&ds^2 =\frac{1}{4}a^2\sec^2 \left[\frac{1}{2}(t'+r')\right]\sec^2\left[\frac{1}{2}(t'-r')\right]\times\nonumber\\
&&\left[-(dt')^2+(dr')^2 +\sin^2 r' d\Omega_2^2\right]\;,
\end{eqnarray} where $-\pi <t'+r' <\pi$, $-\pi <t'-r' <\pi$, $r'\geq 0$. The new
coordinates $r'$ and $t'$ are related to the original coordinates
 $\eta$ and $r$ by
\begin{equation}\label{eta}
\eta=\frac{1}{2}\tan\left[\frac{1}{2}(t'+r')\right]+
\frac{1}{2}\tan\left[\frac{1}{2}(t'-r')\right]
\end{equation}
\begin{equation}\label{r}
r=\frac{1}{2}\tan\left[\frac{1}{2}(t'+r')\right]-
\frac{1}{2}\tan\left[\frac{1}{2}(t'-r')\right]\;.
\end{equation}
Obviously, our flat FRW spacetime filled with a tachyon scalar field and
whose EoS  lies in the range $-1<w<-1/3$
can be mapped into the part of
the Einstein static universe determined by the values taken
by $\eta$ in the interval $-\infty<\eta<0$, which corresponds to the ranges $-\pi<t'<0$ and $0 < r' <\pi$. From these, the resulting Penrose diagram follows after determining the region of the Einstein static space which is conformal to our tachyonic flat space-time.
\begin{figure}
\begin{center}
\includegraphics[width=.8\columnwidth]{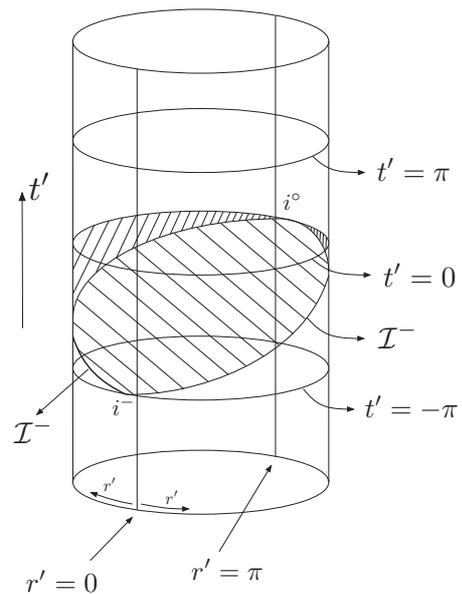}
\caption{\label{fig:1} The flat FRW spacetime filled with a
tachyonic scalar field is conformal to the Einstein static
universe for the EoS range $-1<w<-1/3$. This representation looks similar to
that of the de Sitter space, although it covers a
larger $t'$-interval.}
\end{center}
\end{figure}

\begin{figure}
\begin{center}
\includegraphics[width=1
\columnwidth]{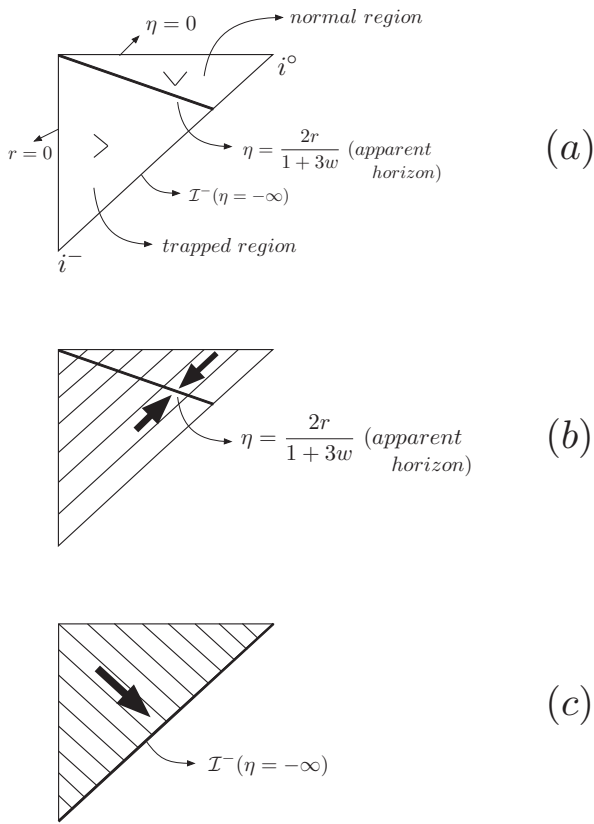}
\caption{\label{fig:2} Penrose diagram of a flat FRW universe filled with a tachyonic scalar field for the range $-1<w<-1/3$. The apparent horizon,
$\eta=2r/(1+3w)$, divides the space-time into a normal
and a trapped region (a). The information contained in the universe can be projected
along future light-cones onto the apparent horizon (b), or
along past light-cones onto past null infinity $\mathcal{I^-}$ (c). Both are preferred screen-hypersurfaces.}
\end{center}
\end{figure}

The parts of the Einstein
static cylinder which are conformal to the tachyonic flat FRW
spacetime for $-1<w<-1/3$ are shown in Fig.\ \ref{fig:1}. The conformal region runs from $t'=0$ to an extreme $t'< 0$. The corresponding Penrose diagram is plotted in Fig.\ \ref{fig:2}.

Now, following the prescription given in \cite{Bousso:2002ju}, we can
construct the holographic screens in our tachyonic flat FRW universe.
The apparent horizon is given by
$\eta=2r/(1+3w)$. The interior of the apparent horizon,  $\eta\geq 2r/(1+3w)$, can be projected along future light-cones centred at $r=0$, or by means of a
space-like projection, onto the apparent horizon.
 The exterior, $\eta\leq 2r/(1+3w)$, can also be
projected by future light cones, but in the opposite direction, onto the apparent horizon. Alternatively, the entire flat tachyonic universe can be projected along past light-cones onto the past null
infinity. The two holographic preferred screens that encode the entire space-time, given by the apparent horizon
$\eta=2r/(1+3w)$ and the past null infinity $\mathcal{I}^-$, are plotted in Fig.\ \ref{fig:2}.

\section{Covariant holography in a tachyonic phantom universe}\label{sec:phtachyonDE}

Phantom dark energy \cite{Caldwell:1999ew,Starobinsky:1999yw} has already confirmed its validity as a dark energy candidate \cite{Carroll:2003st,Singh:2003vx,Cline:2003gs,Sami:2003xv}.
Moreover, Planck latest results \cite{Ade:2013lta} plus WMAP low-\emph{l} polarisation (WP), when combined with Supernova Legacy Survey (SNLS) data, favor the phantom domain ($w<-1$) at 2$\sigma$ level for a constant $w$
\begin{equation}\label{Planck}
w=- 1.13^{+0.13}_{-0.14} \;(95\%; Planck+WP+SNLS)\;,
\end{equation} while the Union2.1 compilation of Type Ia supernovae (SNe Ia) is more consistent
with a cosmological constant ($w=-1$). If we combine Planck+WP with measurements of $H_{0}$ \cite{Riess:2011yx}, we get for a constant $w$
\begin{equation}\label{Planck2}
w=- 1.24^{+0.18}_{-0.19}\;
\end{equation} which is in tension with $w=-1$ at more than the 2$\sigma$ level. Also, for the SNLS3 and the Pan-STARRS1 survey (PS1 SN)
data sets, the combined SNe Ia + Baryon Acoustic Oscillations (BAO) + Planck data
yield a phantom equation of state at $\sim 1.9\sigma$ confidence \cite{Shafer:2013pxa}. The above observational results, in addition to theoretical motivations, are compelling enough to justify the study of the phantom sector in more depth.

The phantom regime, which implies a violation of the dominant energy condition

\begin{equation}
p_{t}+\rho_{t}=\frac{V(\phi)\dot{\phi}^{2}}{\sqrt{1-\dot{\phi}^{2}}}<0\;,
\end{equation} can be obtained by  Wick rotating the tachyon
field so that $\phi\rightarrow i\Phi$, where the field $\Phi$
can be viewed as an axion tachyon field \cite{GonzalezDiaz:2004vq}, as the scale factor
$a(t)$ and the field potential $V(\Phi)$ keep being positive.
In this phantom case, the solution of Eqs.\ (\ref{Fried1}) and (\ref{Fried2}) for the scale factor yields \cite{GonzalezDiaz:2004as}.

\begin{equation}\label{scalephantom}
a(t)=\left[a_0^{3\left(1-|w|\right)/2}+
\frac{3}{2}\left(1-|w|\right)(t-t_{0})\right]^{2/\left[3\left(1-|w|\right)\right]}\;,
\end{equation} which in terms of the conformal time
\begin{equation}\label{confph}
\eta=\int\frac{dt}{a(t)}=\frac{2}{(1-3|w|)a^{(1-3|w|)/2}}
\end{equation} becomes

\begin{equation}\label{scconfph}
a(\eta)=\left[\frac{(1-3|w|)\eta}{2}\right]^{2/(1-3|w|)}\;.
\end{equation}

It is worth noting that in this phantom case $\eta$ runs from $\eta=-2/(3|w|-1)a_{0}^{(3|w|-1)/2} \equiv \eta_{0}<0$ at $t=t_{0}$, to $\eta=0$ at the
 big rip when \begin{equation}\label{tbr}
t \equiv t_{br}=t_{0}+\frac{2}{3(|w|-1)a_0^{3(|w|-1)/2}}=t_{0}-\eta_{0},
\end{equation} to finally reach positive infinity as $t\rightarrow\infty$, therefore the interval is $\eta_{0}<\eta<+\infty$.

The field potential is given by \cite{GonzalezDiaz:2004as}
\begin{equation}
V(\Phi)=\frac{3\sqrt{|w|}}{8\pi
G\left[a_0^{-3(|w|-1)/2}-\frac{3}{2}\sqrt{|w|-
1}(\Phi-\Phi_0)\right]^2}\;,
\end{equation}
with $\Phi_0\rightarrow -i\phi_0$. We note that both this
potential and the phantom tachyon energy density,
\begin{equation}
\rho_{t}=\frac{3}{8\pi G\left[a_0^{-3(|w|-1)/2}
-\frac{3}{2}(|w|-1)t\right]^2}\;,
\end{equation}
increase with time up to blowing up at $t=t_{br}$, to steadily
decrease towards zero thereafter.

However, in order for this description to be applicable also after the big rip barrier at $t=t_{br}$,
 such that the scale factor remains real and positive in that region, not all values of $w$ are allowed
 but only those that satisfy the discretization condition \cite{GonzalezDiaz:2005sh}
\begin{equation}\label{discretew}
w=-\frac{1}{3}\left(1+\frac{2n+3}{n+1}\right),\;\;\; n=0,1,2,...
\end{equation}

Similarly to what we did in Sec.\ 2, we locally obtain the metric (\ref{confmetric}) for the Einstein
static universe, where $\eta$ and $r$ are given by Eqs.\ (\ref{eta}) and
(\ref{r}).

\begin{figure}
\begin{center}
\includegraphics[width=.7\columnwidth]{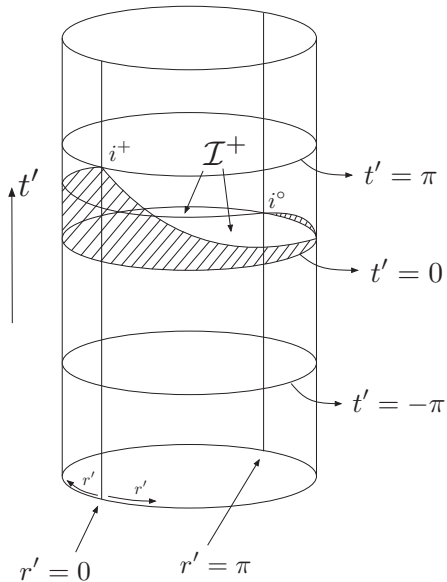}
\caption{\label{fig:3} The flat FRW space-time filled with a
tachyonic phantom energy scalar field is conformal to the Einstein static
universe for the range $-\infty<w<-1$. This flat space looks similar to
that of the de Sitter space ($p=-\rho$), although it covers a
larger $t'$-interval.}
\end{center}
\end{figure}

\begin{figure}
\begin{center}
\includegraphics[width=1.2\columnwidth]{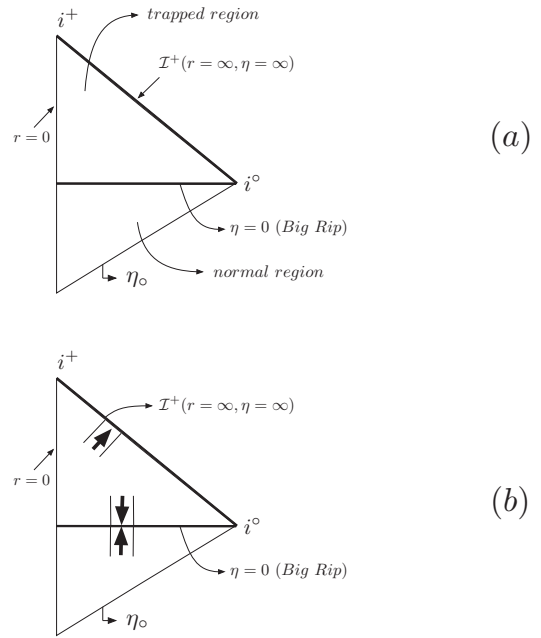}
\caption{\label{fig:4} Penrose diagram of a flat  FRW
universe filled with a tachyon phantom scalar field for
$-\infty<w<-1$. The apparent horizon, also located at
$\eta=2r/(1+3w)$, divides the space-time into a normal
and a trapped region (a). The information contained in the universe can be projected
along future light-cones from the normal region or
along past light-cones from the trapped region, both onto the apparent big rip horizon. It can also be projected along future
 light cones onto the future null infinity $\mathcal{I^+}$ (b). Both the big rip and $\mathcal{I^+}$ are preferred screen-hypersurfaces.}
\end{center}
\end{figure}

Hence, the flat spacetime filled with a phantom tachyonic field we have just considered, which has an
EoS $p=w\rho$ with $-\infty <w<-1$, can also be
mapped into those parts of the cylindric Einstein static universe
which are determined by the values of
conformal time $\eta$ we have discussed above. We can see that the part $\eta_0<\eta<0$ of the whole interval $\eta_0 <\eta< +\infty$
will correspond to a subinterval, which depends on $w$, of the range
$-\pi<t'<0$ and $0<r'<\pi$, and the part $0<\eta<+\infty$
 will correspond to
the range $0<t'<\pi$ and $0<r'<\pi$. This mapping is depicted in Fig.\ \ref{fig:3}
and its resulting Penrose diagram is shown in Fig.\ \ref{fig:4}.

\section{Holographic dark energy models}\label{sec:HDE}
Based on the holographic bound on the entropy \cite{GonzalezDiaz:1983yf,'tHooft:1993gx,Susskind:1994vu} and on the validity of effective
local quantum field theory in a box of size L, Cohen et al
\cite{Cohen:1998zx} suggested a relationship between the ultraviolet
and the infrared cutoffs due to the limit set by the
formation of a black hole. This led Li to propose
a most popular model of holographic dark energy (HDE) \cite{Li:2004rb} that can explain the accelerated expansion of the universe and in which
 the infrared cutoff is taken to be the observer-dependent future event horizon which makes the holographic screen.
This model is in good agreement with observational data \cite{Huang:2004wt,Zhang:2005hs,Chang:2005ph,Ma:2007pd,Li:2009bn,Li:2013dha,Xu:2013mic} but has attracted some criticisms, known as the causality
 and circular logic problems \cite{Kim:2012ik}. Nevertheless, Li has recently proposed a new HDE model with action principle \cite{Li:2012xf} in which
 these problems appear to be no longer present and the evolution of universe only depends on the present state
of universe and the future event horizon cutoff automatically follows from the equations of motion. This
   new HDE also complies well with the most
 recent observational constraints \cite{Li:2012fj}.

 In a flat dark energy dominated FRW universe,  Li's model \cite{Li:2004rb} is based on the
  following relation between
the Hubble parameter $H=\dot{a}/a$ and the size of the future event horizon
$R_h$
\begin{equation}\label{Li}
H^2=\frac{\dot{a}^{2}}{a^{2}}=\frac{8\pi G\rho_{t}}{3}=\frac{c^2}{R_h^2}\;,
\end{equation}
where $R_h=a(t)\int_t
^{\infty}dt'/a(t')$ is the proper size of the future event horizon which plays the role of the holographic screen and $c$ is a numerical
 parameter of order unity which
is related to $w$ by $w=-(1+2/c)/3$.
If we now express the scale factor given by Eq.
(\ref{scfact}) as a function of $c$

\begin{equation}
a=\left[a_0^{(c-1)/c}+\frac{c-1}{c}(t-t_0)\right]^{c/(c-1)}=T(t)^{c/(c-1)}\;,
\end{equation} the
proper size of the future event horizon is then given by
\begin{equation}
R_h=- cT(t)^{c/(c-1)}\left(\left.
T(t')^{-1/(c-1)]}\right|_t^{\infty} \right)\;.
\end{equation}
Obviously, if $w>-1$ (i.e. $c>1$) then $R_h=cT(t)$, which is finite
for finite $t$. Therefore, Li's model is only well defined when $w>-1$
.

In the phantom case of Sec.\ \ref{sec:phtachyonDE}, $w<-1$ (i.e. $c<1$), and the proper size of the future event horizon inexorably becomes
infinity, so we may say that it will vanish for phantom energy.

Since Eq. (\ref{Li}) is not well defined for $c<1$ as it leads to $H=0$, Li argued that holographic phantom models were not viable \cite{Li:2004rb}.
However, in the phantom scenario we should use instead of Eq. (\ref{Li})  the following \cite{GonzalezDiaz:2005sh}

\begin{equation}
H_{{\rm ph}}^2 =\frac{\dot{a}^2}{a^2}=\frac{8\pi
G\rho}{3}=\frac{c^2}{R_{br}^2}\;,
\end{equation} where

\begin{equation}\label{rhph}
R_{br} =  a(t)\int_t^{t_{br}}\frac{dt'}{a(t')}
\end{equation}
is the proper size of the future event horizon for the holographic phantom model, being $t_{br}$ the
time at which the big rip takes place.

For the tachyon phantom model Eq.\ (\ref{rhph}) yields

\begin{equation}\label{Rbr}
R_{br} =  c\left[a_0^{3\left(1-|w|\right)/2}+
\frac{3}{2}\left(1-|w|\right)(t-t_{0})\right]^{2/\left[3\left(1-|w|
\right)\right]}\;.
\end{equation}

We have then seen that Eq. (\ref{Li}) is no longer valid for a covariant holographic description of
an accelerating universe and that the appropriate holographic screens for the covariant specification are the one
at the big rip hypersurface and the one at the future null infinity $\mathcal{I^+}$ (see Sec.\ \ref{sec:phtachyonDE}).

\section{Conclusions}\label{sec:concl}

We have considered the holography of a flat FRW dark energy dominated universe in which the cause of its accelerated expansion is due to the presence of tachyon scalar field with constant EoS $w$. In order to do so, we have applied
a covariant formalism \cite{Bousso:2002ju} and then have compared the results with those obtained by the HDE with the future event horizon as the infrared cut-off \cite{Li:2004rb,Li:2012xf}.

The more general covariant formalism gives rise to two different
holographic preferred screens. In the dark energy case ($w>-1$) these are located at the apparent
horizon $\eta=2r/(1+3w)$ and at the past null
infinity $\mathcal{I^-}$. On the other hand, in the phantom energy scenario ($w<-1$) one is also located at the apparent horizon, which is the big rip hypersurface in this case, and the other at the future null infinity $\mathcal{I^+}$. When we establish the comparison of these results with the ones obtained by using the HDE model \cite{Li:2004rb,Li:2012xf}, whose holographic screen is positioned at the future event horizon, we see that the former allow the definition of fundamental theories based on the existence of a S-matrix at infinite distances, at least when one approaches $\mathcal{I^-}$ or $\mathcal{I^+}$.

There is in addition an apparent contradiction between the implications from the covariant treatment of phantom holography and the fact that phantom energy is characterized by a negative temperature \cite{GonzalezDiaz:2004eu}. We may be led to think that if the preferred holographic screens for phantom energy are located at the big rip and the future null infinity, then the entropy that should be associated with that phantom fluid would be negative definite, implying a definite positive temperature. However, this would be mistaken because the entropy involved in this case is the one defined by the surface area of the future preferred holographic screen, given in this case by  Eq.\ (\ref{Rbr}).

The relevant entropy would actually coincide with the entropy of entanglement \cite{Muller:1995mz} and would be given by

\begin{equation}
S_{Ent}=\left. \alpha R_{br}^2 \right |_{t>t_{br}}
\end{equation} where $\alpha$ is a constant of order unity. In order to calculate the entropy of entanglement we have used the equivalence between the regions before and after the big rip hypersurface. In this case, we have integrated out the region before that surface. This entanglement entropy is definite positive and increases with time, leading again to the conclusion that the temperature of a phantom fluid is definite negative.

\begin{acknowledgements}
The author is grateful to Andrea Maselli for help with the plots. This work was supported by the 'Fundaci\'on Ram\'on Areces' and Ministerio de Econom\'ia y Competitividad (Spain) through project number FIS2012-38816.
\end{acknowledgements}

\end{document}